\documentclass[article]{aa} 
\usepackage{epsfig}
\usepackage{natbib}
\bibpunct{(}{)}{;}{a}{}{,}

\usepackage[utf8]{inputenc}

\usepackage{graphicx}
\usepackage{amsmath}
\usepackage{amsxtra}
\usepackage{amssymb}
\usepackage{mathrsfs}
\usepackage{array}
\usepackage{listings}
\usepackage{soul}
\usepackage{xspace}

\newcommand*{\eg}{e.g.\@\xspace}

\graphicspath{{figures/}}
\begin{document}

\title{Gap and rings carved by vortices in protoplanetary dust}
%subtitle{from numerical simulations to observational signatures }

\author{Pierre Barge\inst{1} \and L. Ricci\inst{2} \and C. L. Carilli\inst{3,4} \and R. Previn-Ratnasingam\inst{1}  }
\institute{
Aix Marseille Universit\'{e}, CNRS, Laboratoire d'Astrophysique de Marseille, UMR 7326, 13388, Marseille, France \\
\email{pierre.barge@lam.fr}
\and  
Harvard-Smithsonian Center for Astrophysics, 60 Garden Street, Cambridge, MA 02138, USA  
\and
National Radio Astronomy Observatory, P. O. Box 0, Socorro, NM 87801, USA
\and
Astrophysics Group, Cavendish Laboratory, JJ Thomson Avenue, Cambridge CB3 0HE, UK }

\date{Version 0.1}

\abstract
  {Large-scale vortices in protoplanetary disks are thought to form and survive for long periods of time. Hence, they can significantly change the global disk evolution and particularly the distribution of the solid particles embedded in the gas, possibly explaining asymmetries and dust concentrations recently observed at sub-millimeter and millimeter wavelengths.}
  {We investigate the spatial distribution of dust grains using a simple model of protoplanetary disk hosted by a giant gaseous vortex. We explore the dependence of the results on grain size and deduce possible consequences and predictions for observations of the dust thermal emission at sub-millimeter and millimeter wavelengths.} 
   {Global 2D simulations with a bi-fluid code are used to follow the evolution of a single population of solid particles aerodynamically coupled to the gas. Possible observational signatures of the dust thermal emission are obtained using simulators of ALMA and ngVLA observations. }
   {We find that a giant vortex not only captures dust grains with Stokes number $St < 1$  but can also affect the distribution of larger grains (with $St \sim 1$) carving a gap associated to a ring composed of incompletely trapped particles.
  %and dust concentrations escaping from the vortex. 
 The results are presented for different particle size and associated to their possible signatures in disk observations.}
   {Gap clearing in the dust spatial distribution could be due to the interaction with a giant gaseous vortex and their associated spiral waves, without the gravitational assistance of a planet. Hence, strong dust concentrations at short sub-mm wavelengths associated with a gap and an irregular ring at longer mm and cm wavelengths could indicate the presence of an unseen gaseous vortex. }

\keywords{protoplanetary disks -- vortices -- accretion -- planetary formation -- submillimeter}

\titlerunning{gap and rings}
\authorrunning{P. Barge et al.}

\maketitle

%%%%%%%%%%%%%%%%%%%%%%%%%%%%%%%%%%%%%%%%%%%%%%%%%%%%%
\section{Introduction}
{The trapping of solid particles in gaseous vortices and its consequences on planet formation is an issue investigated today by many authors. Vortices in protoplanetary disks can be produced by various mechanisms like inverse cascade in two-dimensional geometries or specific 2D/3D instabilities like the Rossby wave or the baroclinic instability.\
Three-dimensional studies have confirmed that such vortices can form and survive in spite of the elliptical instability that predominantly affects the core of the vortices \citep{Lesur2010,Richard2013}. Ultimately, they can be depicted as quasi-2D structures with a nearly stationary evolution. \
In two-dimensional Keplerian flows, vortices are known to capture very effectively the solid particles embedded in the gas \citep{Barge95}. The trapping mechanism is so efficient that particle concentration can be strong enough for
significant back-reaction of the solid particles onto the gas. Dust feedback makes possible the growth of instabilities like Rayleigh-Taylor \citep{chang2010}, Kelvin-Helmoltz \citep{johansen2006}, or streaming \citep{youdin2007}. In vortices, the confinement of the solid particles was recently addressed by \citet{Fu2014, Raettig2015, Crnkovic2015, Baruteau2016, Surville2016}.\\
Here, our paper focuses on the global distribution of the solid particles in a disk hosting a persistent gaseous vortex but does not study the detailed complex evolution inside the vortex core which will be addressed elsewhere. Our goal is to explore how a giant vortex can impact the dust surface density of a protoplanetary disk at a global scale.\\
The most appropriate place for the formation and survival of these vortices are the optically thick regions of the disk where MRI is ineffective due to the weak ionization levels of the gas. For example, the boundary between the dead-zone and the turbulent inner regions has attracted a special interest since it can be the place of a density enhancement with the possible growth of the Rossby wave instability \citep{Lovelace99,Li2001,Varniere2006}. \

A study of these disk outer regions has an additional interest since it can be confronted to circumstellar disk observations, particularly with the new generation of high-resolution instruments. Interferometric observations at sub-millimeter to centimeter wavelengths probe material in the disk with enough angular resolution to highlight asymmetries due to the presence of vortices. Whereas images with high signal-to-noise in spectral line typically probe molecular gas in the disk upper layers because of the high optical depths, imaging of the dust thermal emission allows to investigate the spatial distribution of dust particles in the disk midplane, where vortices are expected to form. Strong azimuthal asymmetries in the dust thermal emission have been observed in a number of young disks at sub-millimeter and millimeter wavelengths, \eg  ~~Oph IRS 48 \citep{Marel2013, Marel2015} 
, LkH$\alpha$ 330 \citep{Isella2013}, HD 142527 \citep{Casassus2013}, SAO 206462, SR 21 \citep{Perez2014}.
Such asymmetries are often attributed to dust trapped in large-scale vortices which are able to collect large amounts of dust grains with sizes of the order of $\sim 1$ mm. Another indirect way to detect vortices in disks is the fortuitous eclipses of a young star by the trapped dust in a highly inclined disk \citep{bar2003, grinin}.\\ 
In this paper, numerical simulations are performed to follow the evolution of dust particles embedded in a gas disk that contains a vortex. The solid particles are aerodynamically coupled to the gas and the presence of the vortex is found to significantly change their global distribution around the star. Depending on their size, the particles are either (i) captured and concentrated inside the vortex \citep{Barge95,Fuente2001,Chavanis2000} or (ii) repelled  from the vortex orbital radius where a gap forms. The carving of a gap in the dust distribution is a forgotten consequence of the dust/vortex interaction that we qualitatively present and discuss in this work. The results of the numerical simulations are presented in the form of surface density maps to make easier confrontations with the observations. They are also illustrated and corroborated by computing the trajectories of passive solid particles in a steady velocity field. Global simulations and trajectory computations were performed for a wide range of particle sizes.

The paper is presented as follows: Section 2 describes the disk/vortex model and the bi-fluid numerical simulations; Section 3 presents the main results of the numerical simulations and the density maps; Section 4 describes the dust vortex interactions explaining the different structures observed in the dust distribution; Section 5 presents simulated observations derived from the density and temperature maps using ALMA and ngVLA simulators. The results of the simulated observations are discussed in Section 6, and the main results of this investigation are summarized in Section 7.\ 

\section{Models and numerical simulations}\label{simu}
To study the evolution of the dust spatial distribution we performed 2D numerical simulations with a bi-fluid (gas$+$dust) code and use a simple model of protoplanetary disk in the following framework:\\
(i) the disk extends from 20 to 100 AU from the central star, a large enough region to get predictions that can be compared with the observations;\\
(ii) the surface density and the temperature of the ``background" structure of the disk, i.e. before a vortex is introduced, are given by simple power-laws: $\Sigma \propto r^{-p}$ and $T \propto r^{-q}$, respectively; \\
(iii) the gas is assumed to be a mixture of hydrogen and helium with a mean molecular weight $\mu =2.34 m_H$ where $m_H$ is the mass of an hydrogen atom; \\
(iv) the dust is described as spherical particles with varying sizes 0.04 mm $< s <$ 2cm and volume density of $\rho_\circ=$ 0.8 g cm$^{-3}$, similar to the dust model considered by \citet{Tazzari2016}; \\
(v) the vortex is introduced at the beginning of the simulation in the form of an approximate solution of the incompressible Euler's equations.\
\subsection{Disk parameters}
The disk model we used discards the standard MMSN assumption and rests on recent observations of protoplanetary disks. We have chosen the primordial disk around the Oph IRS 48 young star as our reference case. This disk shows the strongest asymmetry in the azimuthal distribution of dust particles in a disk observed to date, which could be the signature of mm- and cm-sized dust trapped by a large-scale vortex \citep{Marel2013, Marel2015, Lyra2014, Zhu2014}.\

The required parameters for the disk model are then deduced from the observations of this peculiar object. The analysis of the ALMA observations of the dust continuum emission and CO isotopologues allowed \citet{Bruderer2014} to derive a model for the radial profile of the gas distribution in the regions out of the vortex. \
Although it is not our goal to find a model that reproduces in detail the observations for this disk, we use this case as a fiducial case to investigate the effects of a giant vortex on the global distribution of gas and dust in a real disk system.\

Following \citet{Bruderer2014}, the radial profile of the gas surface density can be described by a power-law function with index $p=1.1$. We consider a gas surface density of 0.32 g cm$^{-2}$ at 60 AU from the star, which is the radial distance of the vortex structure as seen in the Oph IRS 48 disk \citep{Marel2013}. This is a factor of 10 greater than the value for the gas density considered in \citet{Bruderer2014} at the same location in the disk. We chose this larger value because subsequent radiative transfer modeling of disks has shown that masses derived from observations of CO isotopologues can increase by up to an order of magnitude when one properly treats the isotope-selective photodissociation of these molecules \citep{Miotello2014,Miotello2016}.
 
On the other hand, the radial profile of the gas temperature is a power-law with an index of $q = 0.56$, with a value of 100 K at 60 AU (top right panel of Figure 8 in \citet{Bruderer2014}). The vertical scale height of the disk is derived from hydrostatic equilibrium at this temperature and the gravitational potential is for a 2 solar-mass star.\

For the particles, different sizes are considered in the present work. We have chosen grain sizes that range from 0.04 mm to 6.4 mm to predict maps for the dust continuum emission at different wavelengths between $\lambda =$ 0.45 mm and 0.9 cm, which can be probed at high angular resolution and sensitivity with the ALMA and future ngVLA interferometers. Although the observed emission realistically comes from a combination of grains spanning several orders of magnitude in sizes, the most efficient emitting grains have sizes $s \approx \lambda / (2\pi |m|)$, where $m$ is the refractive index of the emitting dust \citep[$m \approx 2 - 3$ for dust in protoplanetary disks, \eg ~][]{Rodmann2006}.

The simulations performed in this paper allow us to explore the dependance of the dust/gas friction on the dust/vortex interaction and its consequence on the dust surface density and thermal emission. The initial distribution of the dust is assumed to trace the distribution of the gas with a dust-to-gas mass ratio equal to 0.001 across the whole disk (simulations performed with a mass ratio of $0.01$ are also presented in Appendix A).\\

\subsection{Vortex model}  \label{vort-mod}
In this work we disregard the problem of vortex formation in a disk which has been addressed elsewhere \citep[\eg ~][]{Richard2013, Barge2016}. Hence, the vortex is introduced at the beginning of the simulations and, in a few rotations, adjusts to a quasi-steady vortex structure that lasts long enough to significantly perturb the motion of the dust particles. 2D vortices are, indeed, quasi-steady gaseous structures that can survive over very long time periods, up to thousands of vortex rotations around the star \citep{Surville2015}. Their lifetime in numerical simulations is shorter when dust is included in the disk due to the gas drag, but the present simulations show that the vortices can survive long enough to significantly change the dust surface density and to imprint observable structures in the disks.\\
For the initial conditions we used a simple model of Rossby vortex developed by \citet {Surville2015}. The control parameters of the vortices were selected to get vortices in the "giant vortex class" in order to mimic the large and strong dust asymmetries observed in a number of disks \citep[\eg ~][]{Casassus2013,Marel2013,Marino2015}. 
In this model the gaseous vortex is centered on ($r_0$, $\theta_0$) in polar coordinates and is described with a pseudo enthalpy  $\mathcal{H}(r,\theta )$ approximated by a two-dimensional gaussian function parametrized by an amplitude $\mathcal{H}_0$ and two widths $\delta r=r_0\chi_r$ and $\delta \theta=\chi_r \chi_\theta$, where $\chi_r$ is the normalized radial extent of the vortex and $\chi_\theta$ is its azimuthal aspect ratio. This function is maximum at the vortex center and smoothly connects to zero at the background level.\\
With this Gaussian approximation we found simple expressions for the velocity field of a gaseous vortex when considered in a frame of reference rotating at the mean angular velocity of the flow $\Omega_o$\
\begin{subequations} \label{vort-field}
\begin{align}
 u  & =  \frac{1}{Ro} \frac{\theta -\theta_0}{\delta \theta^2} ~{{\mathcal{H} }\over {r\Omega_K}}  & \\
 v  & = \left[ -  r\left({ r - {r_0}}\over{\delta r^2}\right) + \frac{(1- \gamma) p + q} {2\gamma} \right]{{\mathcal{H} }\over {r\Omega_K}} 
\end{align}
\end{subequations} 
where $Ro$ is the Rossby number at the vortex center and $\gamma=1.4$ is the adiabatic index of the gas (the second term in the bracket is zero for an isentropic vortex). At the beginning of the simulations the vortex is characterized by $Ro= 0.1$, $\chi_r =0.15 $ and $\chi_\theta =6.9 $. This structure is added on top the equilibrium state of the disk, centered at a distance of 60~AU from the star. The evolution of gas and dust is then provided by a numerical integration of  standard fluid equations (see for example  \citet{Inaba2005}).\

\subsection{Numerical Simulations}
The simulations are performed with a bi-fluid code in which the solid component is considered as a fluid without pressure. This approximation requires that Stokes number and dust-to-gas mass ratio be smaller than one \citep{Garaud2004}. In our problem the two conditions are satisfied except (i) if the particle size is larger than one centimeter or (ii) in the core of the vortex and close to its orbital radius where dust strongly accumulates. The code was previously developed to study the dynamical evolution of protoplanetary disks \citep{Inaba2005, Surville2015}.  It solves the inviscid continuity and Euler's equations vertically integrated over the disk thickness. At equilibrium, the gas rotates at slightly less than the Keplerian velocity, due to the radial pressure gradient, with an angular velocity\\
\begin{equation}
\Omega_o(r) = \Omega_K(r) \left( 1 - \eta \right)^{1\over 2}
\end{equation}
where $\eta = {{p+q}\over \gamma} {M_A^{-2}(r)} $ and $M_A$ is the Mach number; the flow is also assumed compressible and adiabatic but non-homentropic. The solid particles are fully coupled to the gas by aerodynamical drag forces in the Epstein's regime and characterized by a non dimensional parameter or Stokes number\
\begin{equation}
St = {\pi \over 4} {{s \rho_\circ}\over \Sigma(r) }~.
\end{equation}
In a gas disk at equilibrium a solid particle drifts inward to the star due to the pressure gradient; its radial velocity is approximately given by
\begin{equation}
u_{dust} = { {St}\over{1 +St^2}} {\eta r\Omega_K},
\end{equation}
that displays a maximum at $St$=1 \citep[\eg ~][]{Nakagawa1986,Inaba2005}.\\
We start the simulations from a two-component disk model at equilibrium on top of which the vortex solution presented in the above model is added. The simulations are performed in a whole ring around the star, large enough to get significant changes in the radial distribution of the dust.\\
The code was already used to study the formation of vortices by the Rossby-Wave or the baroclinic instability \citep{Richard2013, Barge2016} but also to study the coupled evolution of gas and dust \citep{Inaba2006}. This code, which uses the finite volume method and was first presented in \citet{Inaba2005} can keep high accuracy over large integration periods \citep{Surville2015,Surville2016}. \\
Test simulations show a three step evolution: (i) the approximate vortex-solution adjusts in the flow to a vortex that numerically satisfies the fluid equations in $\sim$10 vortex rotations, (ii) the formed vortex survive in the disk for hundreds of vortex rotations, (iii)  the vortex is progressively destroyed by the back-reaction of the solid particles on a longer time scale.\\
The total duration of the integrations necessary for our simulations was determined from a compromise between the vortex lifetime and the time necessary for the vortex to capture and trap significant amount of solid particles. It was, then, empirically set up to 100 orbital periods of the vortex around the star. \\
The numerical resolution is (2048$\times$4080) in the $(r,\theta)$ directions, sufficient to correctly describe particle captures by the vortex. In the radial direction, the resolution corresponds to 274 cells by scale height at 60~AU from the star. %%%%%%%%%%%%%%%%%%%%
Simulations were performed for different values of the particle size.\\

\section{Main results of the numerical simulations}
The simulations provide the spatial distribution of the dust particles after $100$ rotations of the vortex around the star. The results are discussed as a function of the particle size or the Stokes number estimated at the radial position of the vortex.
\begin{figure*}
%\begin{center} 
%\begin{tabular}{t}
\begin{minipage}[c]{.95\linewidth}
\includegraphics[width=5.8cm]{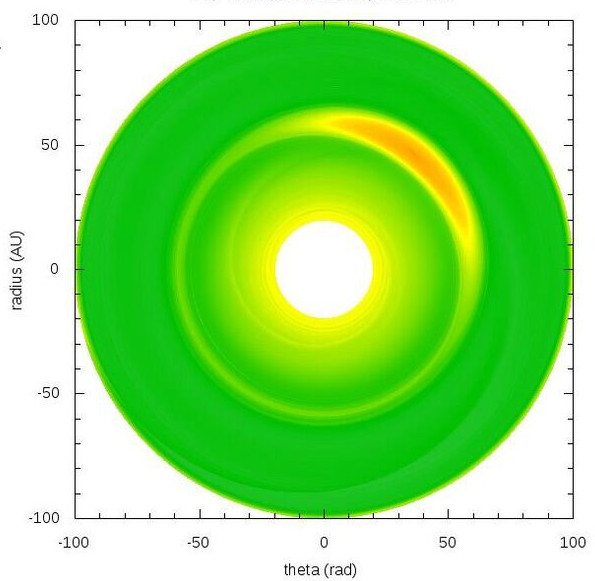} 
\includegraphics[width=5.8cm]{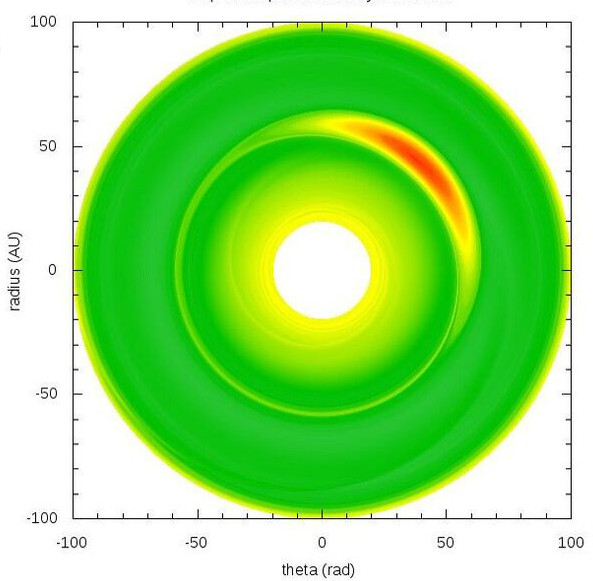} 
\includegraphics[width=5.8cm]{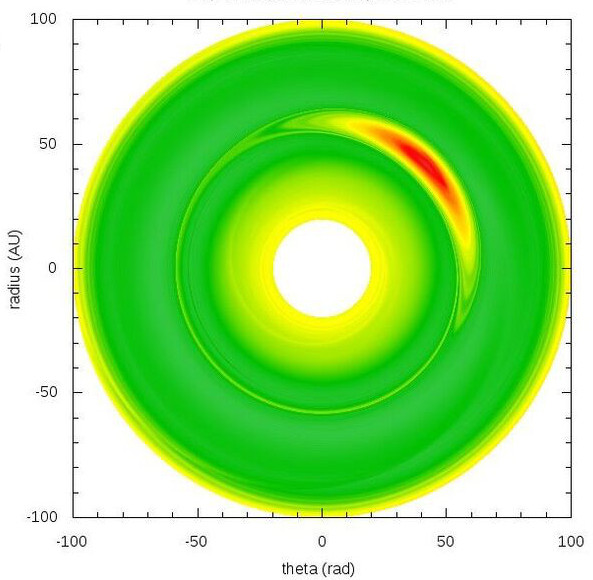}\
\includegraphics[width=5.8cm]{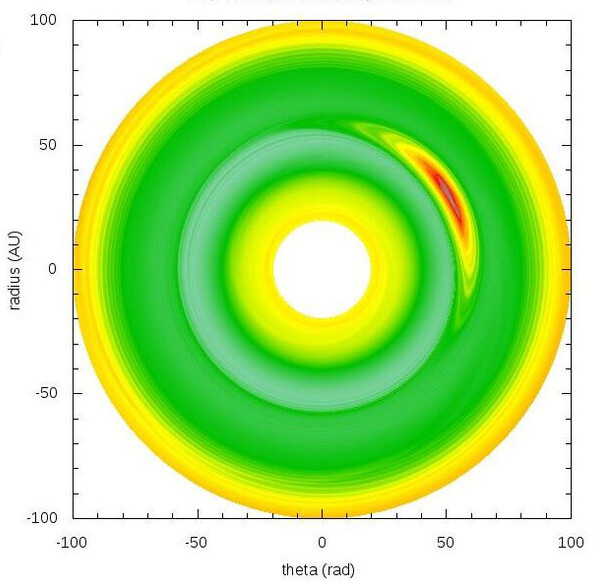} 
\includegraphics[width=5.8cm]{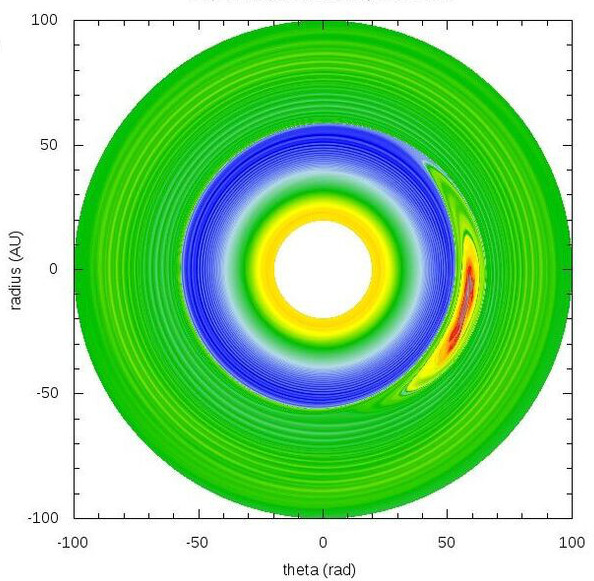} 
\includegraphics[width=5.8cm]{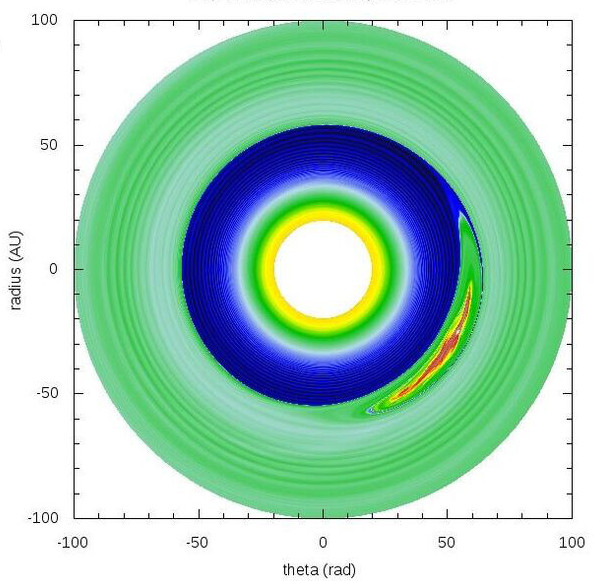}\
\includegraphics[width=5.8cm]{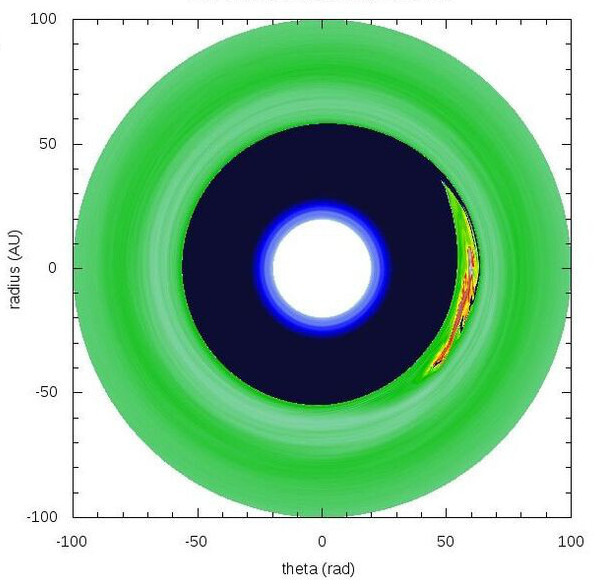} 
\includegraphics[width=5.8cm]{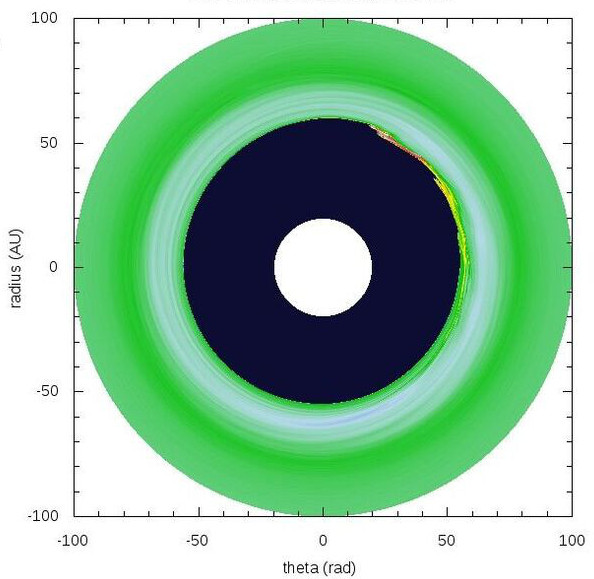}
\includegraphics[width=5.8cm]{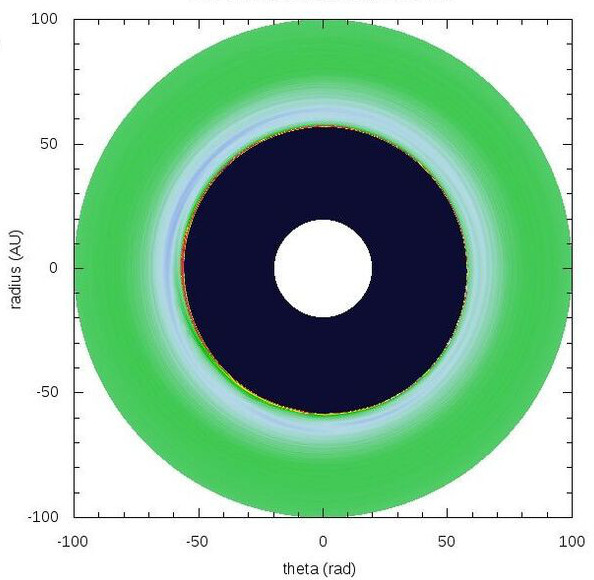}\
\end{minipage}
\begin{minipage}[c]{0.0003\linewidth}
\includegraphics[width=1.2cm,height=3cm]{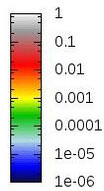}
\end{minipage}
\caption{Dust density in a disk containing a giant gaseous vortex that has completed 100 rotations around the star (units are in $gcm^{-2}$). The vortex is initially located at 60 AU from the star; in each plot the solid particles have a single size equal to: 0.046 mm, 0.092 mm, 0.14 mm, 0.32 mm, 0.74 mm, 1 mm, 2 mm, 6.4 mm, 2 cm, from left to right and from top to bottom, respectively. }
 \label{map}
%\end{center}
\end{figure*}
\subsection{Global distribution of the dust}  \label{result}
Figure \ref{map} shows the density map of the dust for different values of the particle size on which we note a number of striking features.\\
(1) Sub-millimeter sized particles ($St \lesssim 0.063$) concentrate in a regularly-shaped region centered on the vortex core (as commonly expected for passive particles).\\
(2) Millimeter sized particles ($0.063\lesssim St \sim 1$) concentrate in an irregular and inhomogeneous core-region (containing dust knots) while a gap is carved just inside the vortex orbital radius (gap depth increasing with particle size).\\ %%%%%%%%%
(3) Centimeter size particles ($ 1\lesssim St$) are strongly depleted inside the vortex orbital-radius but concentrate in a narrow and knotted structure (arc or ring) where the dust-to-gas mass ratio can overcome unity (see Figs. {\ref{zoom2}} and {\ref{zoom3}}). \\
In this last case we deliberately transgress some of our initial assumptions to explore weaker coupling with the gas avoiding complex developments or the use of a different code.\\
Yet, we note a piling up of the solid particles at the inner boundary of the simulated ring due to the decrease of the drift velocity with the distance to the star; a similar effect occurs at the outer boundary for particle sizes of 0.14 mm and 0.32 mm but it could result from a numerical artefact.
\subsection{Evolution of the dust particles}
Our simulations also provide a step by step description of the dynamical evolution of the dust particles. Three different cases have been distinguished following the values of the Stokes number.\\
\\
$\bullet$ ``Light'' particles ($St < 1$)\\
These particles are directly captured by the vortex and confined for a long time in its core where the dust-to-gas ratio can overcome unity allowing a significant back-reaction of the particles onto the gas. Figure \ref{zoom1} shows the concentration of millimeter-sized particles in the core of the vortex with the formation of a high-density knot that becomes turbulent under dust/gas instabilities. 
After about 40 rotations the dust surface density has a strong jump at the orbital radius of the vortex and the formed gap presents a layered structure. Ridges correspond to a spiral lane of non-captured particles that are slowly drifting inward toward the star.\\
The density contours of the gas are plotted on the same figure and indicate (i) the position of the vortex (concentric elliptical paths) and (ii) the position of density waves associated to the vortex (oblique rays at the top and bottom of the figure). The contours are noisy but their large scale deformations, after $100$ rotations, reveal vortex stretching and turbulent motions} due to the back-reaction of the particles onto the gas.\\
\begin{figure}[h]
\begin{minipage}{1.03\linewidth}
\centering
\includegraphics[width=4.5cm]{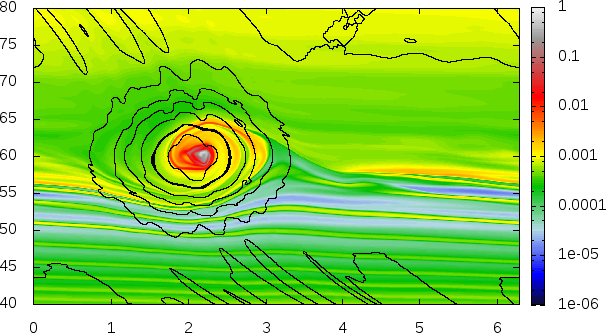}
\includegraphics[width=4.5cm]{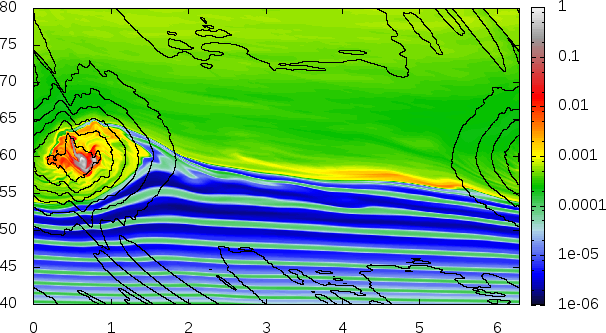}\\
\vspace{0.2cm}
\includegraphics[width=9.cm]{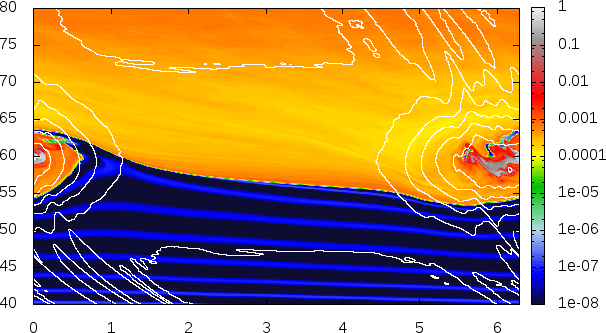}
\end{minipage}
\caption{Dust-to-gas ratio for light particles of size $s=$ 2 mm. The map is plotted in the $(r,\theta)$ plane and the evolution is shown after $20$, $40$ and $100$ rotations of the vortex, from left to right and from top to bottom, respectively. Units of x- and y-axes are AU and radians, respectively.}
\label{zoom1}       % Give a unique label
\end{figure}\\
$\bullet$ ``Optimal'' particles ($St \sim 1$)\\
The capture efficiency is the most effective for particles with $St\simeq 1$. Such particles rapidly concentrate in a small region  of the vortex core (Figure \ref{zoom2}), which is turbulent due the back-reaction of the particles. In the inner part of the disk the gap is deeper than for lighter particles and also presents a layered structure.\\ In this case, the back reaction of the particles onto the gas is strong enough to significantly change the vortex shape after $100$ rotations.\\
\begin{figure}[h]
\begin{minipage}{1.03\linewidth}
\centering
\includegraphics[width=4.5cm]{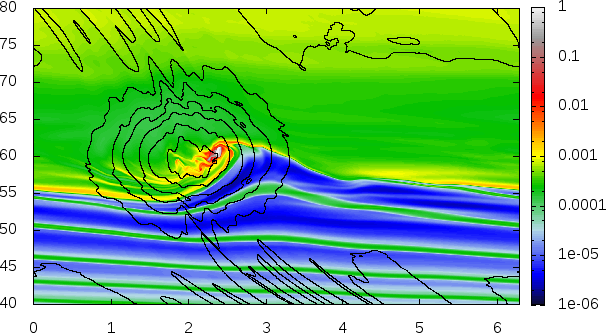}
\includegraphics[width=4.5cm]{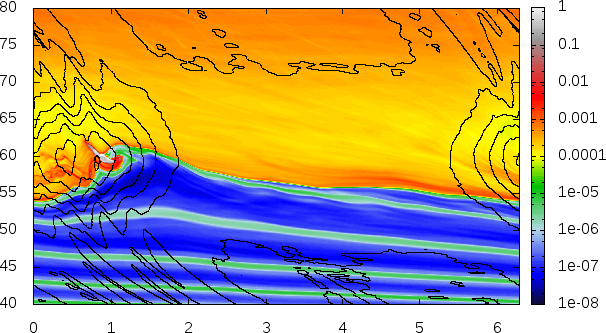}\\
\vspace{0.2cm}
\includegraphics[width=9.cm]{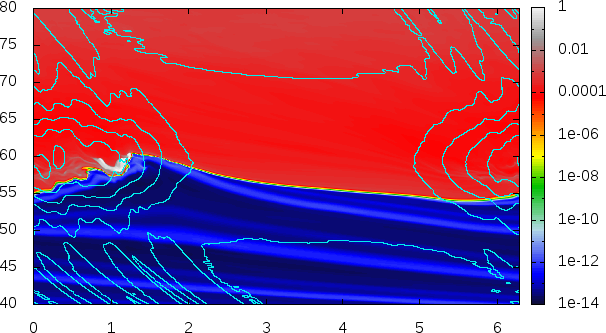}\
\end{minipage}
\caption{Dust-to-gas ratio for nearly optimal particles after $20$, $40$ and $100$ rotations of the vortex. The particle size is $s=$ 6.4 mm. Layout is the same as in Fig. \ref{zoom1}. }
\label{zoom2}       % Give a unique label
\end{figure}\\
$\bullet$ {``Over-optimal" particles} ($St \gtrsim 1$)\\
Particles in the centimeter range marginally violate our initial assumptions but permit to explore dust/gas evolution for weaker coupling. The vortex shape is less affected than for smaller particles  (cf. Fig. \ref{zoom3}). The particles focus in a very small region of the disk that looks like a knotted arc close to the orbital radius of the vortex. This structure evolves in an annular chain of high-density knots which individually covers a surface of the order of $1\times4 AU^2$ or 25$\times$40 numerical cells.  Similar concentrations of particles  were also observed in simulations performed by \citet{Surville2016} when $St\sim 0.1$. Inside this radius the surface density decreases dramatically like in the previous case with also the formation of a gap and its layered structure.\\
\begin{figure}[h]
\begin{minipage}{1.03\linewidth}
\centering
\includegraphics[width=4.5cm]{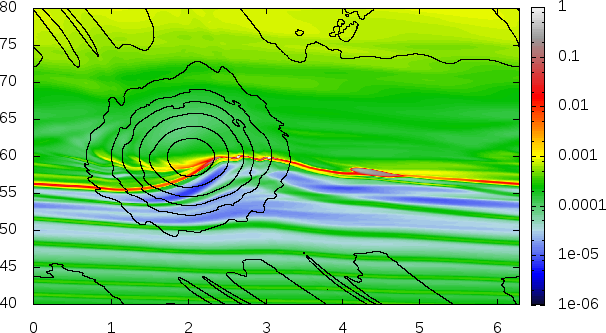}
\includegraphics[width=4.5cm]{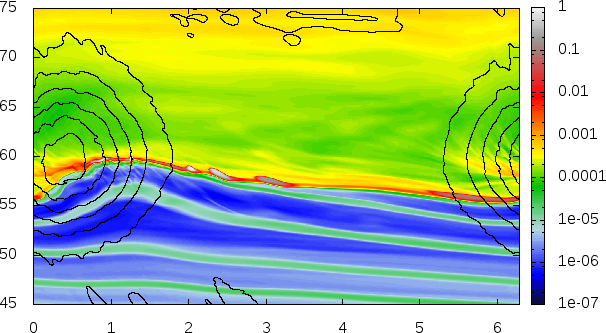}\\
\vspace{0.2cm}
\includegraphics[width=9.cm]{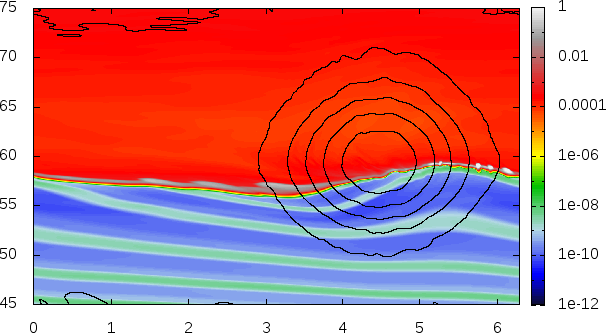}\
\end{minipage}
\caption{Dust-to-gas ratio for a particle size of $s=$ 2 cm after $20$, $40$ and $100$ rotations of the vortex; same layout as in Fig. \ref{zoom1}. }
\label{zoom3}       % Give a unique label
\end{figure}
\section{Dust/vortex interaction}
Our simulations show that the evolution of the solid particles and their distribution around the star strongly depends on the value of the Stokes number $St$. Particles are found either to be trapped and concentrated in the vortex or to form a high-density ring lining a strong density jump at the orbital-radius of the vortex. The capture by a vortex is a well known mechanism due to the local action of the Coriolis force. On the other hand, the carving of a gap with a strong density jump is a new output of the dust/vortex interaction that have to be explained. Here, we propose a number of carving mechanisms.\

\subsection{Capture and radial drift}
\begin{figure}[h]
\begin{minipage}{.91\linewidth}
\centering
\includegraphics[width=4.02cm]{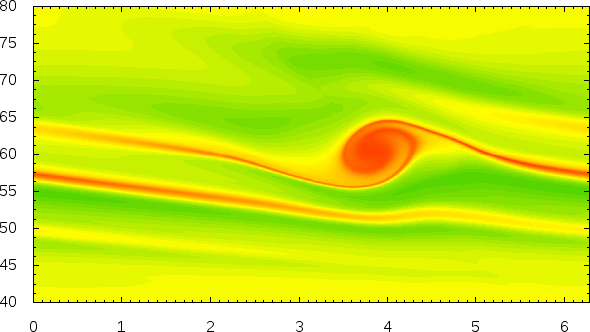}
\includegraphics[width=4.02cm]{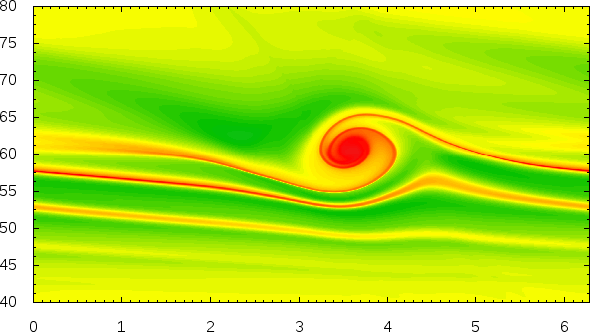}
\includegraphics[width=4.02cm]{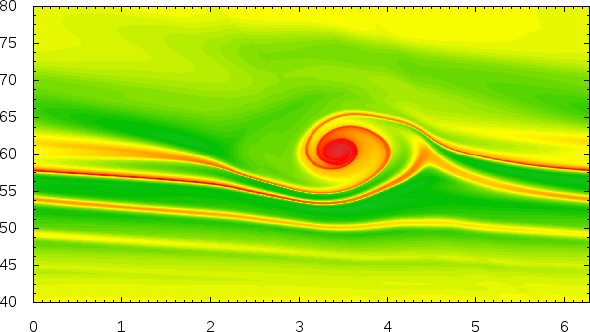}
\includegraphics[width=4.02cm]{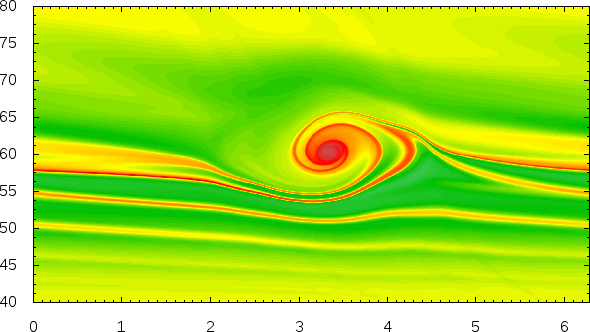}\
\end{minipage}
\begin{minipage}{0.08\linewidth}
\includegraphics[width=0.8cm,height=4.2cm]{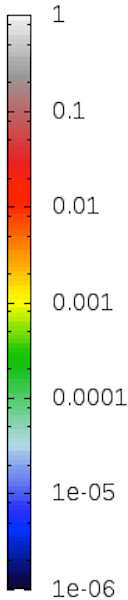}
\end{minipage}
\caption{A dust lane in the vortex's wake that spirals around the star and is progressively captured by the vortex. The dust to gas ratio for particles with size $s=$ 2 mm is plotted after 5, 7, 8, and 9 rotations of the vortex, from left to right and from top to bottom, respectively.}
\label{wake1}       % Give a unique label
\end{figure}

In its motion around the star the vortex captures dust particles and concentrates them in its core. In a shearing sheet context particles are selected when their impact parameter is smaller than a threshold. There are also non-captured particles among which two classes can be distinguished: (i) outward particles that were too far from the vortex orbital radius to be captured, (ii) inner particles that were too slow to catch the vortex.\\
In fact, due to the systematic inward drift, particles of the outward class have clearly a chance to be captured after a second turn of the vortex around the star; this obviously increases the capture efficiency of the vortex during its lifetime. \\
Our simulations show that particles of the inward class are not inevitably lost into the star but have a chance to be captured during additional turns of the vortex around the star. 
This mechanism concerns the particles that were not rapidly trapped by the vortex and are tracing a spiral wake in the disk. \\
Figure {\ref{wake1}} shows the various steps of this mechanism with a dust lane that distorts gradually when approaching the vortex and that finally rolls up around the vortex core. This secondary trapping is observed all along the vortex lifetime and systematically feeds the vortex with particles of the inner disk. \\

As a result, the systematic drift of the solid particles toward the star has a major role in the carving of a gap in the dust distribution. On the one hand, it contributes to a strong depletion of the inner disk due, both, to the systematic drift toward the star and to the secondary trapping by the vortex. By contrast, in the outer disk, the depletion by the vortex can be balanced by the inward flux of solid particles that are crossing the outer boundary of the computational box. Indeed, in our simulations the outer boundary must be considered as a permeable frontier for external particles since they are also drifting inward toward the star. \
\subsection{Collective drag }
At a given stage, dust trapping becomes strong enough that collective effects and back-reaction on the gas are coming into play; so, the dust concentration is dragged away and other concentrations are forming. They are in turn pulled away from the vortex and tend to form a knotted ring (see Fig. \ref{zoom3}).\\
\subsection{Trajectories of the dust particles} \label{traject}
The dust surface density in the presence of a vortex can be also studied in term of the dynamical evolution of dust particles. A simple approach of the problem is to use a vortex model as described in section  (\ref{vort-mod}). In a frame of reference rotating at the angular velocity of the flow, the dynamical evolution of passive solid particles can be described by the motion equations:\
\begin{subequations} \label{motion}
\begin{align}
 \ddot{r}  - r \dot {\theta}^2  ~~~ & = -{ {\Omega_K}\over{St} }( \dot{r} -u)  + 2\Omega_o r \dot{\theta} + \Omega_o^2 r - { {GM_*} \over{r^2} }   \\ 
  r \ddot{\theta} + 2 \dot{r} \dot{\theta}  & = -{ {\Omega_K} \over{St} } (r \dot{\theta} -v)  - 2 \Omega_o \dot{r},  
\end{align}
\end{subequations}
in which $u$ and $v$ are the two components of the gas velocity changes due to the presence in the flow of a persistent structure (vortex and/or spiral waves). These two motion equations are then solved numerically using a second-order Runge-Kutta method.
\subsubsection{With the vortex model}
\begin{figure}[h]
\begin{minipage}{1.0\linewidth}
\centering
\includegraphics[width=4.4cm]{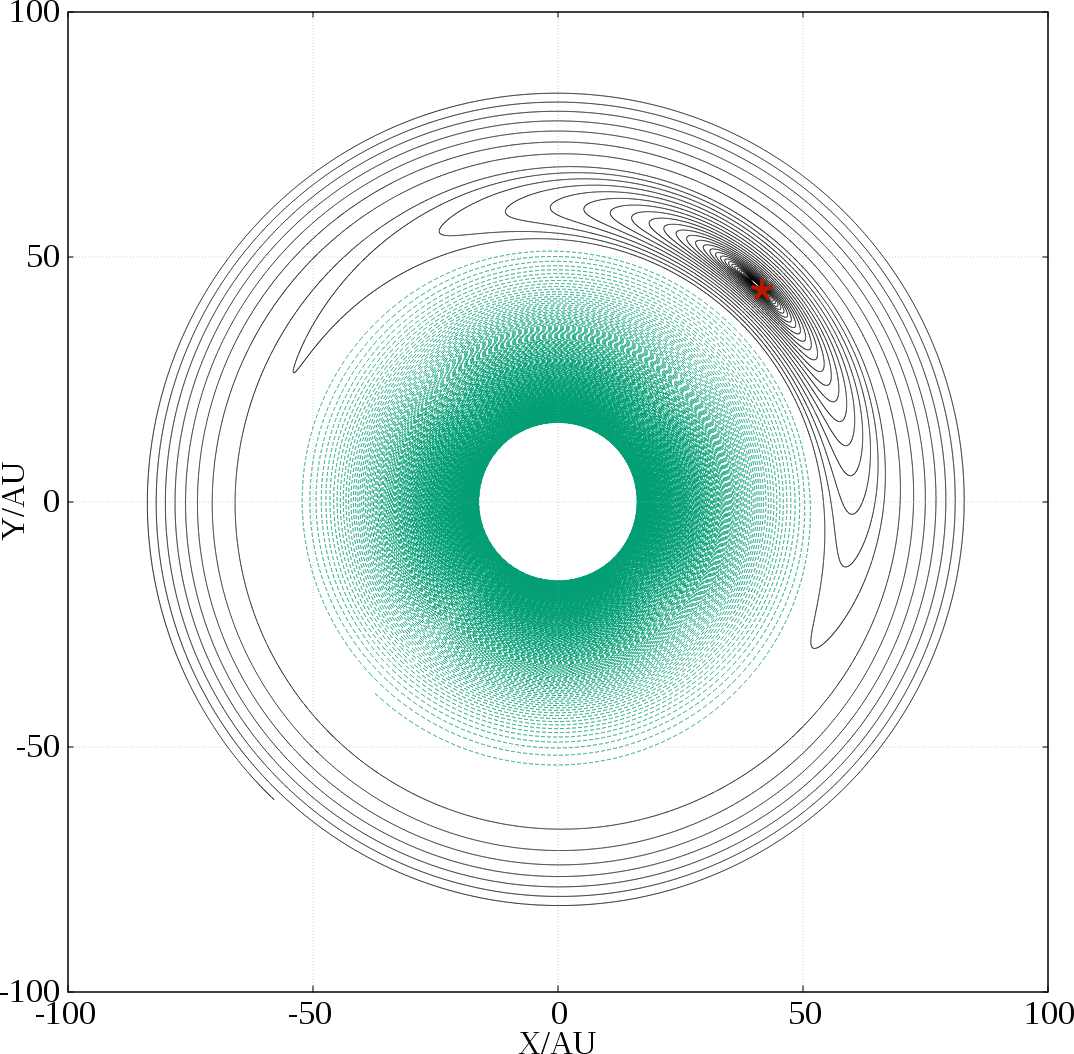}
\includegraphics[width=4.4cm]{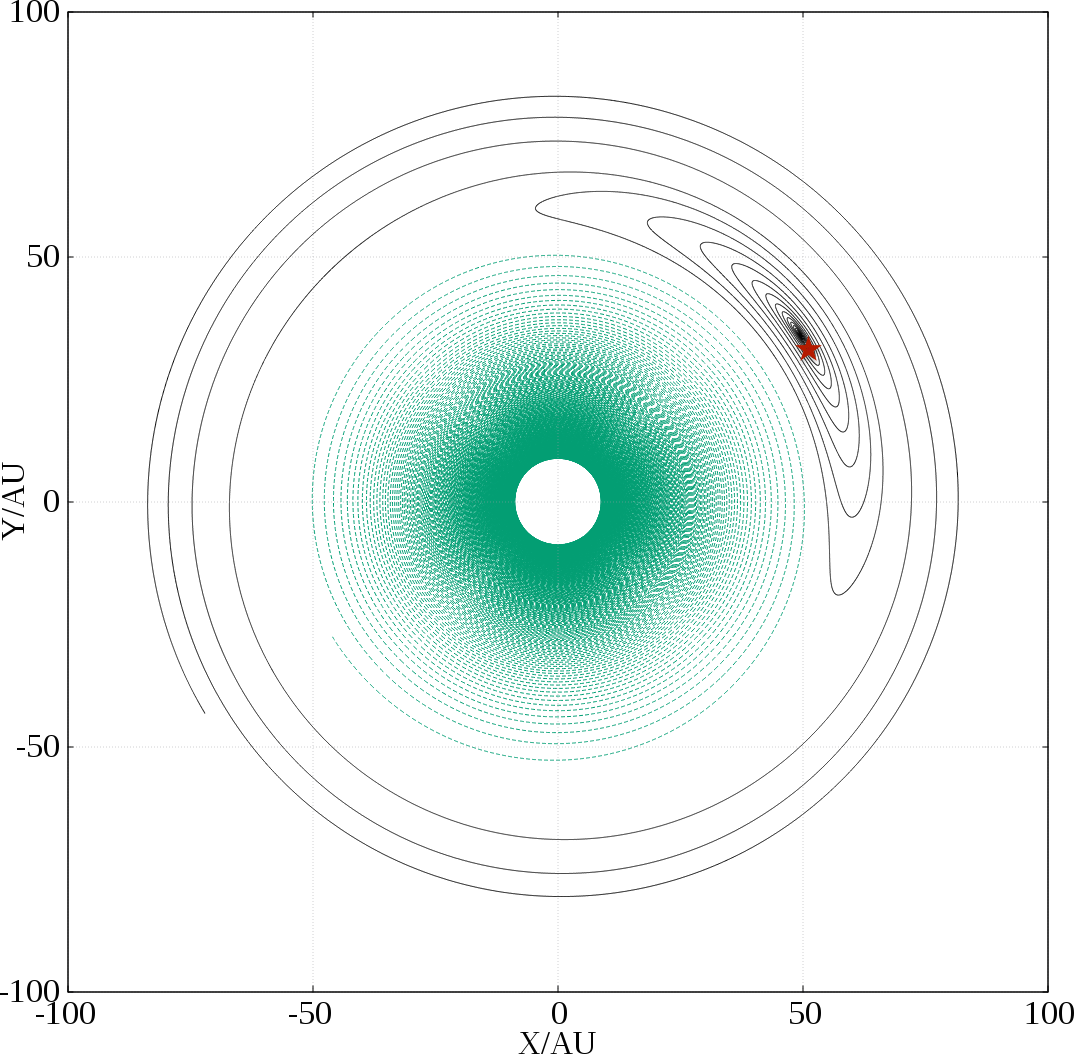}
\end{minipage}
\caption{Trajectories of solid particles interacting with a gaseous vortex. The particle sizes are equal to 0.14 mm and 0.32 mm, from left to right, respectively.}
\label{traj-1}       % Give a unique label
\end{figure}

Particle trajectories were first computed in the vortex velocity-field described in Eqs. {\ref{vort-field} } (with no other contribution) and are plotted in Fig. {\ref{traj-1}}. Two types of trajectories can be distinguished following the value of the impact parameter: (i) vortex trapping (black line) or (ii) inward drift toward the star (green line), with a steep change between the two. Testing various particle sizes we also noted that the transition between trapping and drift is sharper and sharper for increasing sizes; the threshold is located close the core orbital radius in accordance with the results of the numerical simulations. 
\\
The capture trajectories are similar to those computed with approximate vortex models  \citep[\eg ][]{Fuente2001}. 

\subsubsection{With a spiral wave model}
Vortices in Keplerian disks are always escorted by spiral compressional waves they themselves excite in the gas flow. These waves are rotating around the star at the vortex angular-velocity and are also associated to velocity changes of the gas at the crossing of the spiral arms. Due to their coupling with the gas, particles have their motions perturbed at spiral-arm crossing and can exchange angular momentum with the waves, accelerating or decelerating depending on whether their orbit is located outside or inside the core orbital radius, respectively. The importance of this effect depends on the stopping time of the particles but also on the width, amplitude and winding of the waves.
\begin{figure}[h]
\begin{minipage}{.91\linewidth}
\centering
\includegraphics[width=8.5cm]{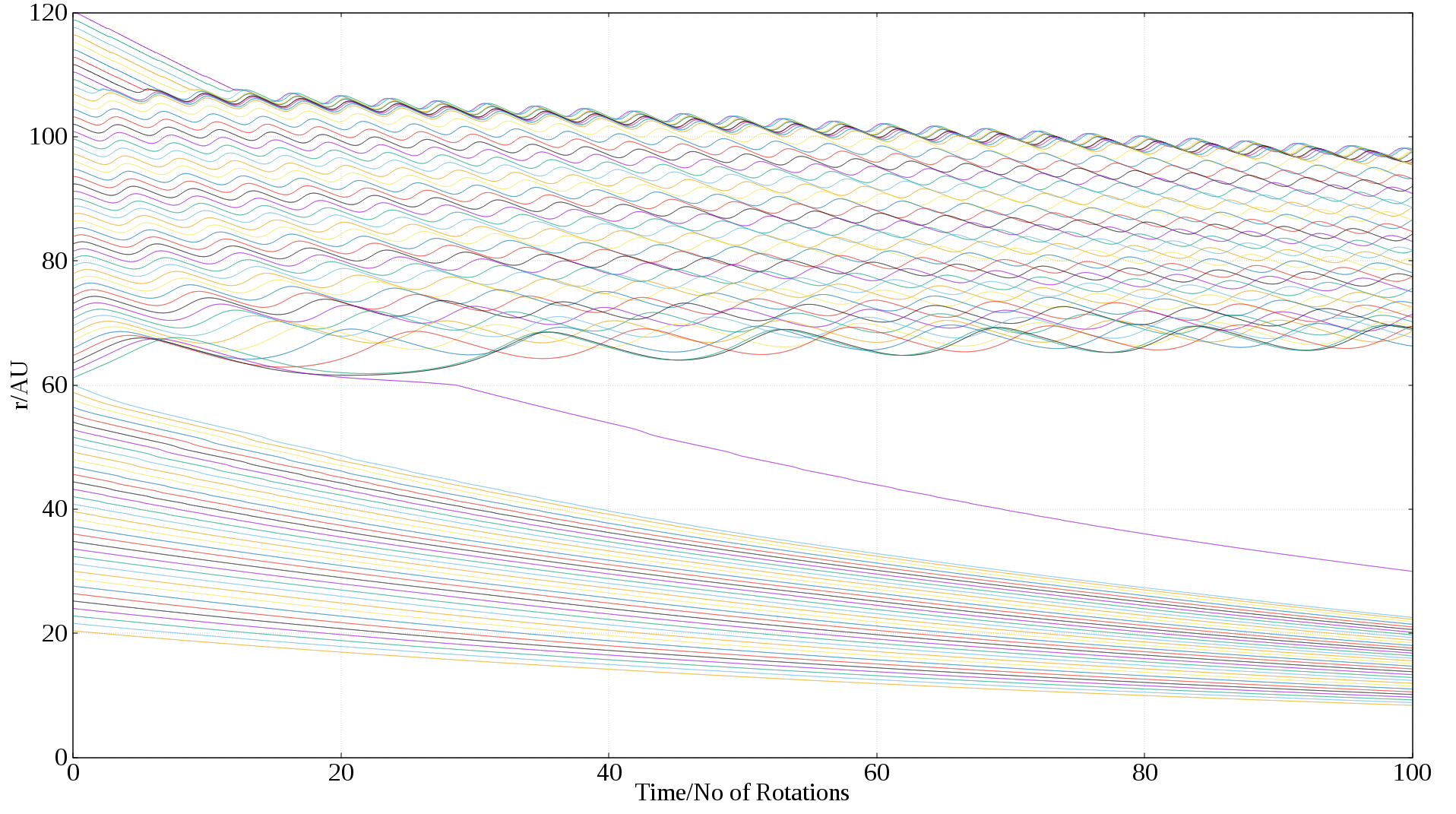}\
\includegraphics[width=8.5cm]{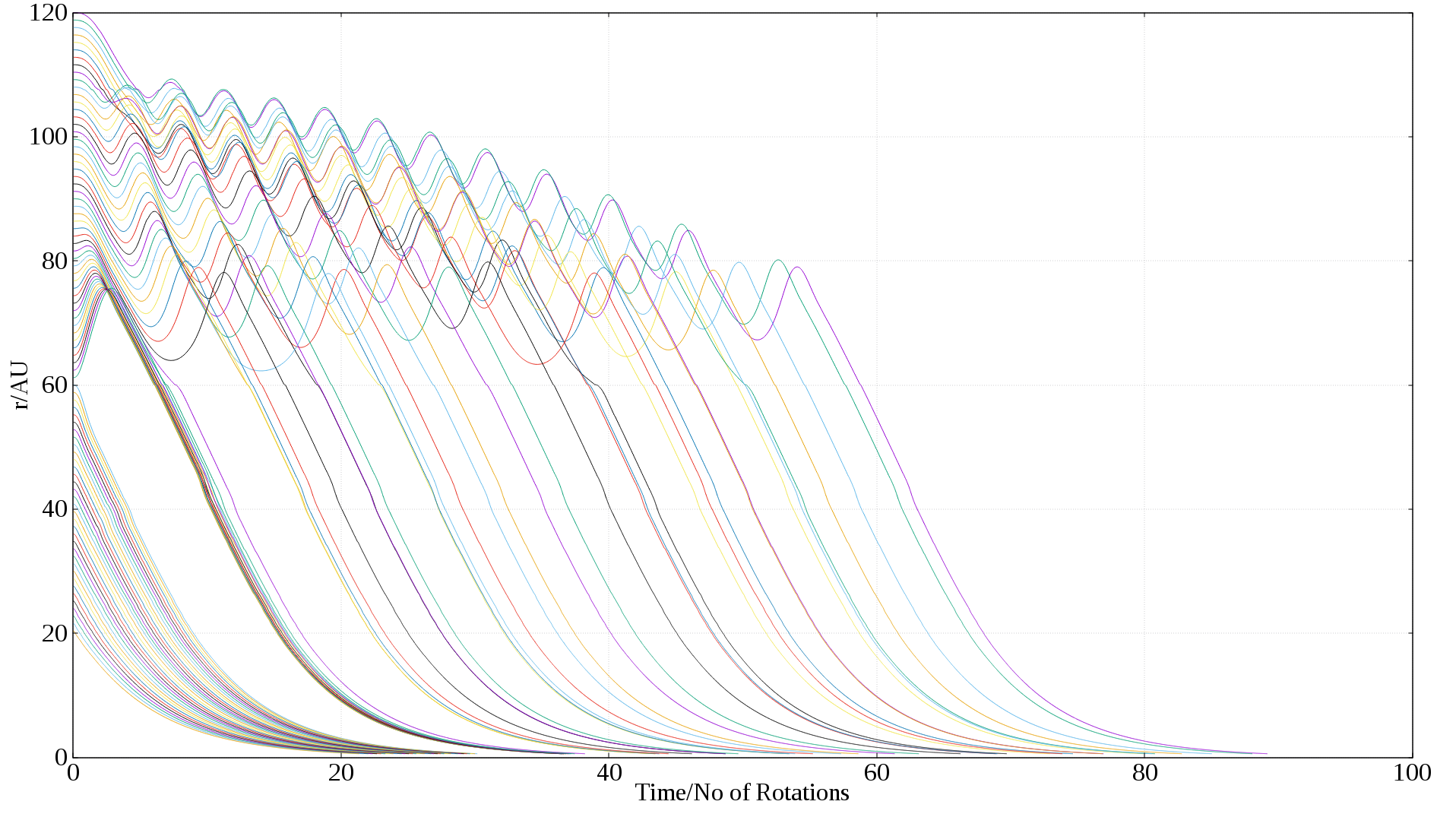} 
\end{minipage}
\caption{Radial drift of solid particles interacting with spiral+vortex (time is in number of vortex rotation). Particle size is $0.32mm$ and $6.4mm$ , from top to bottom, respectively. The various curves are for different values of the impact parameter. }
\label{dec}       % Give a unique label
\end{figure}\\
A detailed study of particle/wave interactions is out of the scope of the present work and our goal, here, is only to capture the main consequences of spiral waves on the dynamical evolution of dust particles coupled to the gas by a drag force. To this end we have taken benefit of the above computations replacing the vortex model by a crude model of a spiral pattern. 
In our toy model the effect of the spiral wave reduces to perturbations of the gas velocity that we have taken in the form
\begin{equation}
\label{spir-field}
\begin{aligned}
	u & =  \epsilon_u ~ f(\theta)  \\
	v & =  [\epsilon_v + r(\Omega_K - \Omega_o)] ~f(\theta),
\end{aligned}
\end{equation} 
where $\epsilon_u$ and $\epsilon_v$ are the amplitude of the velocity perturbations in the radial and azimuthal directions, respectively, and $f(\theta)$ is an arbitrary cut-off function\
\begin{equation}
f(\theta) =  exp[ -( { {\theta -\theta_s}\over {\delta \theta_s} } )^2]
\end{equation}
with $\theta_s \propto (r/r_0 -1)^2 $ to mimic the spiral shape of the perturbation.\\
Then, the motion equations (Eqs. {\ref{motion}}) are solved using the same method as in the above section. The results are reported in Fig. {\ref{dec}} for various values of the impact parameter. \\
The inward drift of the solid particles presents a sharp transition at the orbital radius of the vortex: the drift is slowed down outward while it is accelerated inward. The difference between the two drift rates could help to explain the formation of a plateau lined by a strong gap.\\

\section{Simulation of the observations}\label{simu-obs}
In this section we discuss the observability of the structures predicted by the numerical simulations presented in the previous sections.
We use the results of the numerical simulations for the different grain sizes to investigate the change in the appearance of the dust thermal emission at different wavelengths, as continuum observations at longer wavelengths are typically more sensitive to emission from larger grains~\citep[\eg ][]{Perez2012}.
For the smaller grains of our simulations ($s < 1$ mm) we simulate observations with ALMA, which provides high angular resolution and sensitivity needed to trace the structures discussed in this work.  
For tracing larger grains, which require observations at longer wavelengths, we simulate observations with the future ngVLA interferometer which will achieve high angular resolution and sensitivity at wavelengths longer than $\sim 3$ mm \citep{Carilli2015}, and will be particularly suitable for observations of large grains in protoplanetary disks \citep{Isella2015} \.
The surface brightness map of the dust thermal emission for our models was derived from the dust surface densities and temperatures provided by the numerical simulations. 
We set the absolute scale of each surface brightness map by requiring the total flux to be consistent with the measured sub-mm/mm spectral Energy Distribution of the Oph IRS 48 disk \citep{Marel2015}, which essentially fixes the values for the dust opacity, assumed constant throughout the disk, at each wavelength discussed here.
The disk is seen with a face-on geometry.\\
\begin{figure*}
\begin{minipage}{.98\linewidth}
\centering
\includegraphics[width=6.cm]{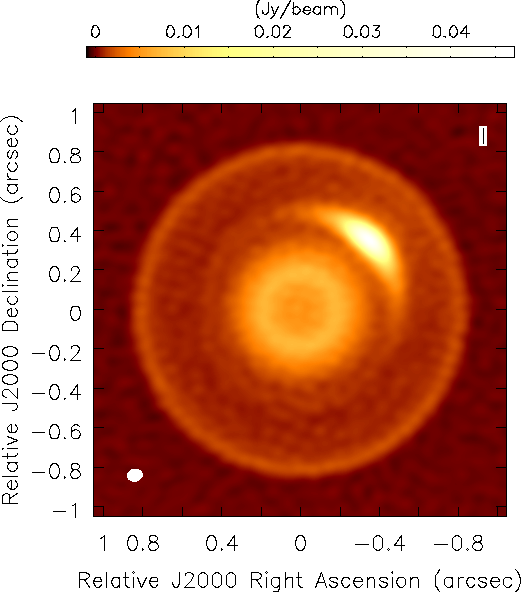} 
\includegraphics[width=6.cm]{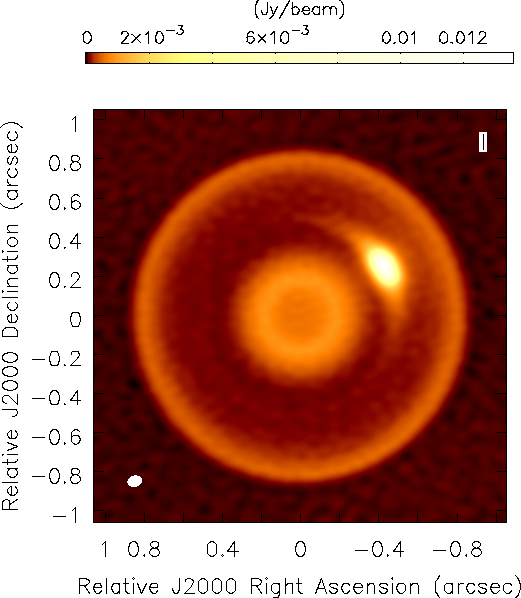} 
\includegraphics[width=6.cm]{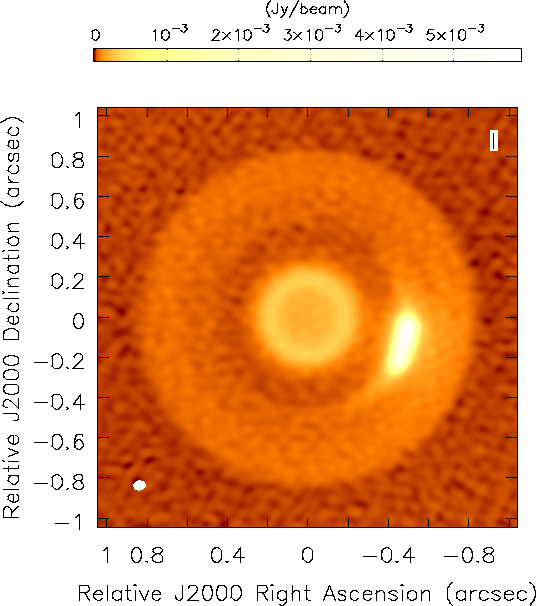} 
\includegraphics[width=6.cm]{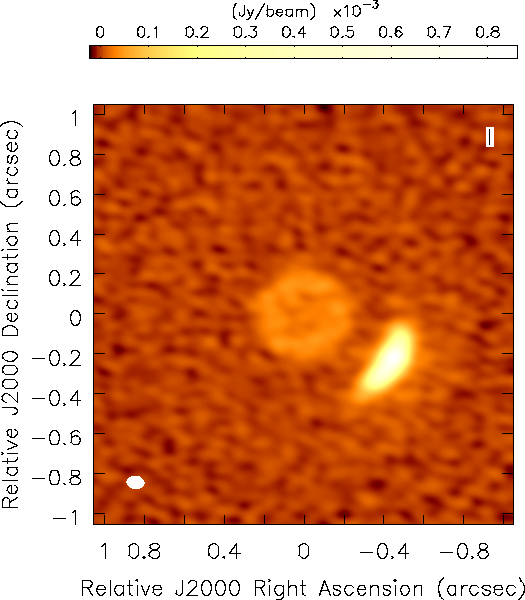} 
\end{minipage} 
\caption{Simulated ALMA images computed from the disk simulations illustrated in Fig. {\ref{map}}. Each map corresponds to a single particle size which is equal to 0.14 mm (wavelength of the simulated observations of 0.43 mm), 0.32 mm ($\lambda =$ 0.87 mm), 0.74 mm ($\lambda =$ 1.3 mm), and 1 mm ($\lambda =$ 2.7 mm), from left to right and from top to bottom, respectively.}
\label{alma1}
\end{figure*}
\subsection{ALMA observations}
We used the online ALMA Observation Support Tool\footnote{http://almaost.jb.man.ac.uk.} (version 4.0) to simulate realistic ALMA observations for a subset of our simulations. 
For our models with grain sizes of 0.14, 0.32, 0.74 mm, and 1 mm, we simulated ALMA observations at wavelengths of 0.43 (ALMA Band 9), 0.87 (Band 7), 1.3 mm (Band 6), 2.7 mm (Band 3), respectively. We considered array configurations that provide an angular resolution of about 80 milliarcsec, or about 10 AU at the distance of Oph IRS 48, for all the ALMA simulations. We found that this angular resolution offers a good trade-off between angular resolution and surface brightness sensitivity for the structures of interest for this paper. An observing time of 5 hours was assumed for each observation. Precipitable water vapor (PWV) levels of 0.5, 1.3 and 1.8 mm were adopted for the simulated observations at wavelengths of 0.43, 0.87 and 1.3 mm, respectively. Figure \ref{alma1} shows the maps returned by the ALMA Observation Support Tool after performing deconvolution with the CASA CLEAN algorithm (Briggs weighting with robust parameter of 0.5).\
\subsection{ngVLA observations}
\begin{figure*}
\begin{minipage}{.90\linewidth}
%\begin{center}
\centering
\includegraphics[width=5.5cm]{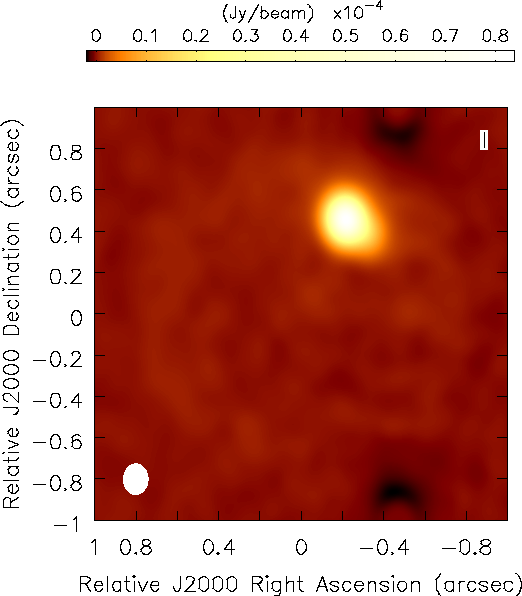} 
\includegraphics[width=5.5cm]{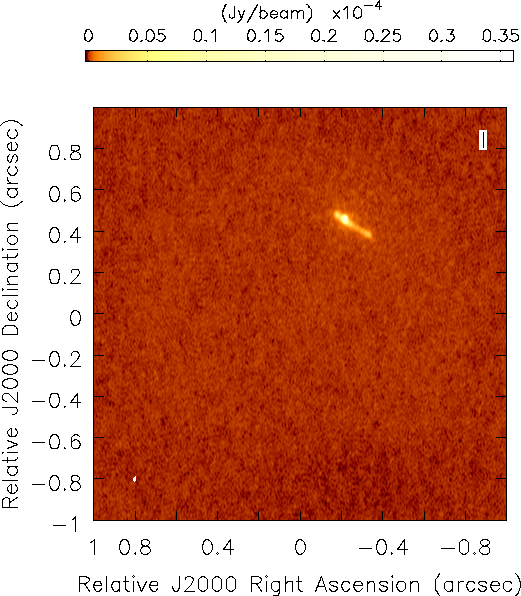} 
\includegraphics[width=5.5cm]{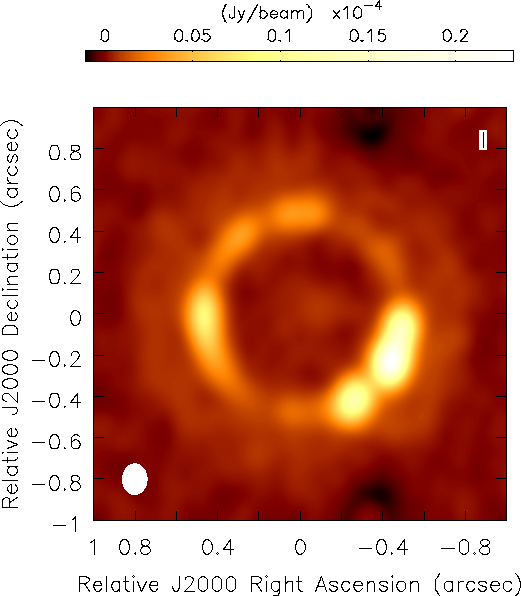} 
\includegraphics[width=5.5cm]{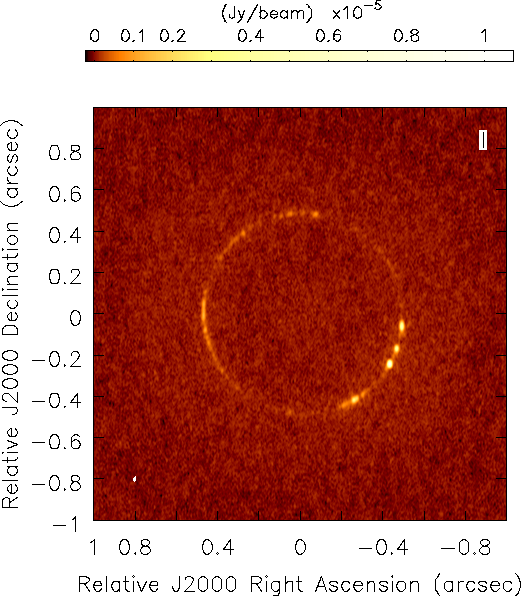} 
\end{minipage}
\caption{Simulated ngVLA images computed from the disk simulations illustrated in Fig.\ref{map}. Each map corresponds to a single particle size which is equal to 6.4 mm ($\lambda=$ 3.7 mm) and 2.0 cm  ($\lambda=$ 1.0 cm), in top and bottom rows, respectively. The images on the left panels were obtained after tapering the interferometric visibilities with a 0.1 arcsec beam; the images on the right panels fully exploit the maximum angular resolution expected for the ngVLA at these wavelengths, as described in the text.}
%\end{center}
\label{ngvla1}
\end{figure*}

For observations at longer wavelengths tracing larger grains, we considered the ngVLA interferometer, which is currently being designed as a future large radio facility operating at wavelengths between about 3 mm and 30 cm (Carilli et al. 2015). We simulated ngVLA observations at wavelengths of 3.7 mm (frequency of 80 GHz) and 1.0 cm (30 GHz) for our models with grain sizes of 6.4 mm and 2.0 cm, respectively. \\
For the ngVLA simulations we used the SIMOBSERVE task in CASA adopting an array configuration made of 300 antennas with longest baselines of 300 km and an inner core to recover the emission from the largest scales of the disk.  Thermal noise was added using the CASA tool SETNOISE and was scaled to provide a final rms noise of about 0.11 and 0.040 $\mu$Jy beam$^{-1}$ in the simulated images at 3.7 mm and 1.0 cm, respectively, and for the highest angular resolution at these frequency, i.e. 5 and 12 mas. These noise values would correspond to a total integration time of about 100 hours with the current ngVLA specifications. Deconvolution was performed using CLEAN with Briggs weighting with a robust parameter of 0.5. The obtained images are shown in Fig. \ref{ngvla1} \footnote{More detailed information on these ngVLA simulations can be found here: \texttt{http:library.nrao.edu/ public/ memos/ ngvla/ NGVLA-11.pdf}.}.\\
%%%%%%%%%%%%%%%%
\section{Discussion of the simulated observations}
We studied the evolution of solid particles in a protoplanetary disk that hosts a persistent gaseous vortex and listed the various structures formed in the global dust distribution. One of our main goals was to identify the possible observational signatures of these structures to better assess the presence of a gaseous vortex in the disk. To this end we performed numerical simulations of a protoplanetary disk using $\rho$ Oph IRS 48 as a reference, as this disk is suspected to host a large scale gaseous vortex.\\ 
We found that, depending on their size (or Stokes number), the solid particles (i) either are captured and confined in the vortex forming a high density region in the core where dust/gas instabilities can grow, (ii) or form high density dust blobs that align in a knotted structure (irregular arc or ring) close to the orbital radius of the vortex core. \\ 
Inside this radius the dust surface density decreases dramatically forming a very sharp gap.\\
Simulated images produced with the ALMA and the future ngVLA simulators show that a number of these large scale structures could be observed.

\begin{figure*}
\begin{minipage}[c]{0.95\linewidth}
\centering
\includegraphics[width=13.cm]{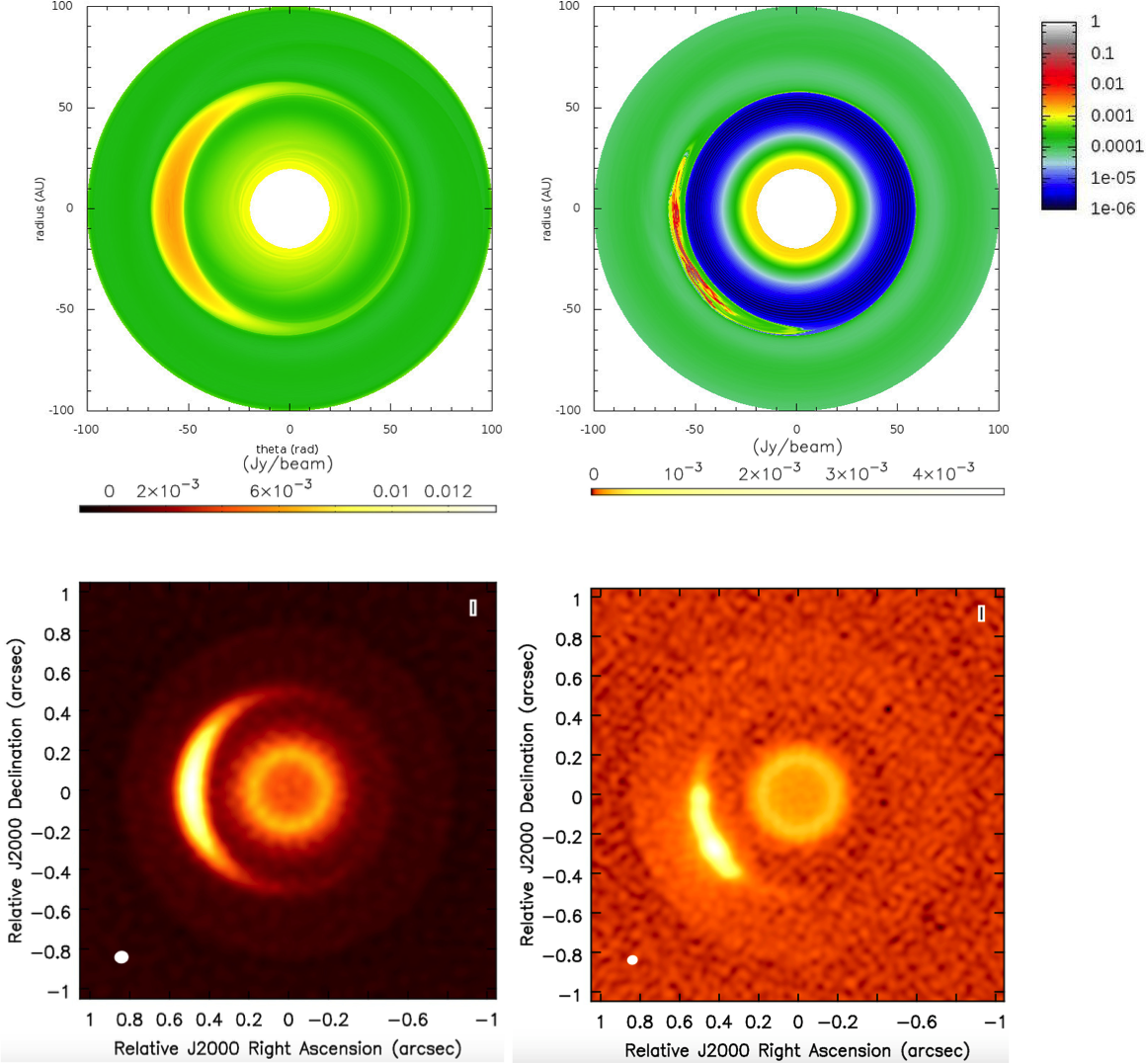}
%%\fbox{
%%\begin{minipage}[t]{.65\linewidth}
%%\centering
%%\includegraphics[width=5.8cm]{../Figures-2016/0,0046cm-asp12-2.jpg}  \hspace{0.1cm}
%%\includegraphics[width=5.8cm]{../Figures-2016/0,0740cm-asp12-2.jpg} 
%%\includegraphics[width=5.90cm]{../Figures-2016/alma_00046_asp12_80mas_5h_0,45mm.png}
%%\includegraphics[width=5.90cm]{../Figures-2016/alma_0074cm_ok.png} \hspace{0.2cm}
%%\end{minipage}
%\begin{minipage}[b]{0.07\linewidth}
%%\includegraphics[width=1.3cm,height=3.2cm]{../Figures-2016/Echelle.jpg} \vspace{2.1cm} %\hspace{0.0cm}
%\end{minipage} 
\end{minipage}
\caption{Dust density and ALMA images simulated in a disk containing a vortex with an aspect ratio of 12. Top, the density maps correspond to a single population of solid particles with size equal to  0.046mm and 0.74mm , from left to right, respectively. Bottom, simulated ALMA images at a wavelength of 0.45mm and 1.3mm, from left to right, respectively.}
\label{alma2}
\end{figure*}  
The most striking features in the simulated ALMA images are: (i)  the high density region in the vortex core is clearly observable at all the wavelengths investigated here; (ii) the depletion of solid particles around the vortex and the piling up at the inner and outer boundaries of the disk are particularly visible at $\lambda$ = 0.43 and 0.87 mm, respectively; (iii) the sharp density jump at the orbital radius of the vortex is distinguishable mostly at $\lambda$ = 1.3 mm, whereas at $\lambda$ = 2.7 mm only the bright dust emission from particles trapped by the vortex and from the inner disk is detected, as due to the lower ALMA sensitivity to dust emission at longer wavelengths (Fig. \ref{alma1}).\\
At each of the investigated ALMA observing bands, the peak of the surface brightness map is always found at the location of the dust trapped by the vortex. This indicates that only observations at with enough sensitivity and angular resolution can highlight the presence of substructures due to the interaction between the vortex and the disk. \\
Furthermore, the azimuthal variation of the location of trapped particles with different sizes is clearly seen from the maps at different wavelengths. 
Figure \ref{alma2} shows the density maps and the simulated ALMA images obtained in the case of a vortex with an aspect ratio two times larger than in the standard case studied above. The azimuthal asymmetries produced in the simulated images have clearly a larger azimuthal extent that could better mimic true observations.
\\ 
At longer wavelengths, the high resolution and sensitivity observations achievable with the ngVLA have the potential to probe the high density regions in the vortex as well as the knotted ring structures expected for pebbles at $\lambda$ = 3.7 mm and $\lambda$ = 1 cm (Fig. \ref{ngvla1}).\\

\section{Summary}
We have shown the results of 2D simulations obtained with a gas/dust code to calculate the spatial distribution of solid particles with different sizes as expected in a disk hosting a giant gas vortex. Our simulations show that a giant vortex not only captures dust grains with Stokes number $St < 1$, but can also affect significantly the distribution of larger grains ($St \gtrsim1$) by carving a gap as well as producing a narrow ring-like distribution of pebbles due to dust concentrations escaping from the vortex core. 
We used the results of our code to calculate the expected surface brightness map at different sub-mm to cm wavelengths, tracing emission from solids with different sizes. We then use ALMA and ngVLA simulators to test the observability of the substructures predicted by our code using high angular resolution and sensitivity observations in dust continuum with these sub-mm and radio interferometers. 
Most of the substructures predicted in this work could be observed for a disk with properties similar to those derived for the $\rho$ Oph IRS 48 disk. However, this is not the case with the presently available observations of this object.
\\
%%%%%%%%%%%%%%%%%%%%%%%%%%%%%%%%%%%%%%%%%%%%%%%%%%%%%
\begin{acknowledgements}
Computations were performed using HPC resources from GENCI [TGCC and CINES] (Grant -  x2016047407) and also at LAM on the mib MPI cluster maintained by CESAM. We want also to thank the anonymous referee for his useful comments.
\end{acknowledgements}

\bibliographystyle{aa}
\bibliography{giant}

\begin{appendix}
\section{Stronger dust-to-gas ratio}\label{strong} 
\begin{figure}[h]
\begin{minipage}{1.03\linewidth}
\centering
\includegraphics[width=4.5cm]{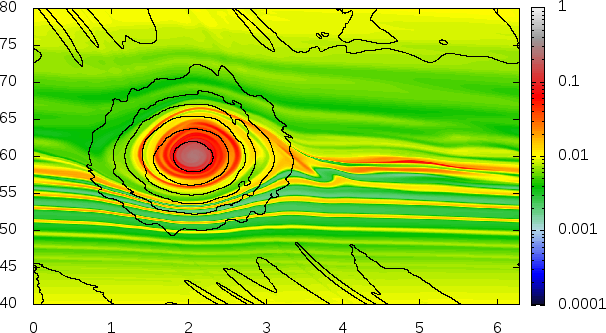}
\includegraphics[width=4.5cm]{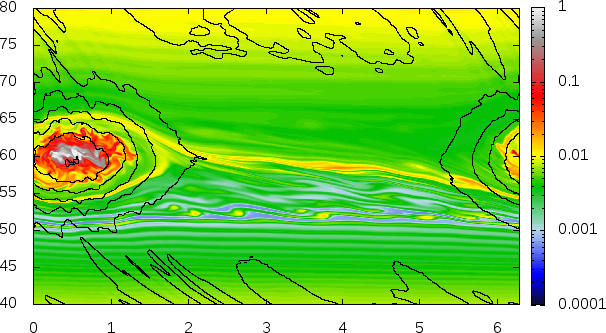}\\
\vspace{0.2cm}
\includegraphics[width=9.cm]{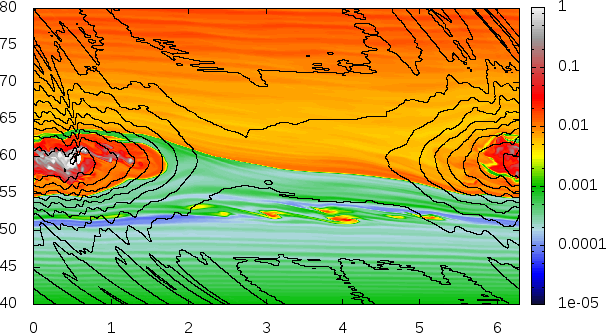}\
\end{minipage}
\caption{Dust to gas ratio for nearly optimal particles after $20$, $40$ and $100$ rotations of the vortex. The particle size is $s=6.4mm$. Layout is the same as in Fig. \ref{zoom1}. }
\label{zoom4}       % Give a unique label
\end{figure}

Simulations starting with a dust-to-gas ratio equals to $0.01$ ((i.e) the cosmic abundance) were also performed in the same context as in section \ref{simu}. To put all the heavy material of a disk in a single population of solid particles is, of course, an unrealistic assumption that was considered only as a limiting case. Moreover this is not a favorable assumption to maintain a vortex against friction drag and dust/gas instabilities. Fig. {\ref{zoom4}} shows an example of such simulations when the particle size is equal to $6.4mm$. The evolution of the dust distribution is similar to the one observed in Fig. \ref{zoom3} with shorter time scales and different numerical values. We estimated it does not make sense to simulate observations in this case. 
\\

\end{appendix}

\end{document}